\documentclass[aps,pra,superscriptaddress,showpacs,twocolumn]{revtex4-2}
\usepackage{braket}
\usepackage{graphicx}
\usepackage{amsmath,amsbsy,amsfonts,amssymb}
\usepackage{epsfig}
\usepackage{hyperref}
\usepackage{xcolor}
\usepackage{ulem}
\renewcommand{\vec}{\mathbf}

\usepackage{verbatim}
\usepackage{xcolor}
\definecolor{lightgrey}{rgb}{0.8, 0.8, 0.8}
\definecolor{remark}{rgb}{0.0, 0.5, 0.69}
\definecolor{Fritz}{rgb}{0.2,0.6,0.2}

\newcommand{\changeS}[1]{\textcolor{blue}{#1}}

\newcommand{\remark}[1]{\textcolor{remark}{#1}}
\newcommand{\remarkF}[1]{\textcolor{Fritz}{#1}}
\begin{document}

\title{Einstein's basement - A kinematic sector complementing special relativity} 
\author{Fritz Riehle}
%\email[e-mail: ]{sebastian.ulbricht@ptb.de}
\affiliation{Physikalisch-Technische Bundesanstalt (PTB), Bundesallee 100, 38116 Braunschweig, Germany}
%\email[e-mail: ]{fritz.riehle@ptb.de}
\author{Sebastian Ulbricht}
%\email[e-mail: ]{sebastian.ulbricht@ptb.de}
\affiliation{Physikalisch-Technische Bundesanstalt (PTB), Bundesallee 100, 38116 Braunschweig, Germany}
\affiliation{Fundamentale Physik für Metrologie FPM, Physikalisch-Technische Bundesanstalt (PTB), Bundesallee 100, 38116 Braunschweig, Germany}
\affiliation{Institut für Mathematische Physik, Technische Universität Braunschweig, Mendelssohnstraße 3, 38106 Braunschweig, Germany}
%\date{\today}

\begin{abstract}
We revisit the concept of particles as it is used in special relativity. 
The presented model
treats the energy-momentum relation of relativistic particles as the upper branch of a generalized energy-momentum relation of quasi particles.
These particles emerge
from a forbidden crossing between the constant energy of a massive particle in rest and the linear energy momentum relation of a massless particle. 
The lower branch, a regime dubbed as Einstein's basement, gives rise to particles with different kinematics that is analyzed in the low-velocity limit.  Allowing for gravitational interaction between those particles we find both attraction and repulsion, depending on their velocity with respect to an absolute space. This absolute frame is only relevant for Einstein's basement and does not affect the relativistic dynamics of regular matter as long as no coupling between the branches is considered, while weak coupling induces local Lorentz violations. 
We finally discuss whether our approach can be used to model phenomena of dark matter and dark energy.

\end{abstract}
\maketitle
%\noindent Corresponding author: \textit{fritz.riehle@ptb.de} 

\section{Introduction}
In classical mechanics, the concept of a {\it particle} has proven extremely useful to abstract the motion of isolated bodies. This concept of particles has the advantage that almost everybody has a clear intuition about its meaning.
However, the interpretation of what a particle is becomes more complicated when it goes to the physics of the smallest scales described within the framework of quantum mechanics and ultimately by quantum field theory \cite{zee10}. There, ``particles'' are the excitations of quantum fields.  

This picture holds not only for fundamental fields, and the corresponding fundamental particles, such as electrons and photons, but also can be used to describe emergent phenomena, e.g. in solid state physics, where the excitations of non-fundamental fields are treated as {\it quasi particles}.
Fundamental particles differ from quasi particles like phonons or polaritons \cite{kit05, hua51, hua51a} in the fact that no underlying system they emerge from, such as an atomic lattice or a polarizable optical medium, is known.
However, history tells us that the term ``fundamental'' always shifts to a more basic layer when our understanding of the world and our insight into the underlying mechanisms increases. 
While once the atom was considered as indivisible, in the last century one discovered protons and neutrons and even found that they are compounds of quarks and gluons as the particles of quantum chromodynamics \cite{gre07d}, considered as fundamental to date. 

Quasi particles, as well as fundamental particles have a well defined energy momentum relationship (EMR), emerging from underlying symmetry principles.
A prominent example is the standard model of particles \cite{she12r} where the irreducible representations of the Lorentz symmetry group give rise to the zoo of particles, respecting the EMR of special relativity \cite{sch17r}.
While being the most efficient description of nature at small scales we have today, the standard model still is capable to explain a number of observations \cite{lyk10}.
Second, the dynamics of regular matter in galaxies hints on the presence of additional, dark matter \cite{and22,ber18,ber16c} built up by yet unknown particles.

One way to tackle these problems might be to relax the restrictions on what is considered to be a ``fundamental'' particle and, instead, to assume such particles to be quasi particles of a yet unknown underlying system.

Following this line, our paper is organized as follows. In Sec.~\ref{sec:the_model}, we apply 
the no-level crossing theorem to the two conceptually different types of particles, namely those with constant rest energy and those with purely kinetic energy. 
Below the well known EMR of special relativity we obtain a second one, marking a region that is dubbed henceforth as ``Einstein's basement''.
We then discuss the formal equivalence of the derived energy momentum  relationships with the ones \cite{kit05,hua51,hua51a} encountered in the polariton system (Sec.~\ref{sec: Polaritons}).
In Sec.~\ref{sec:new_dynamics} 
we apply classical mechanics to the new EMR and derive and discuss the corresponding equations of motion analogously to the mechanics of special relativity and Newtonian physics.
We then study two-particle systems of gravitationally interacting quasi particles from both regions and discuss consequences of our findings and possible extensions of our model in Sec.~\ref{sec:conclusion}.

Focus of this work is 
to consider the dynamics of classical point particles with a modified dispersion relationship in comparison to Newtonian mechanics. 
These studies lay the foundation to investigate further concepts like spin or gauge invariance that could be considered when fields and their dynamics are constructed ``on top'' of the given energy-momentum relationship, and are quantized afterwards.

\section{The model}
\label{sec:the_model}
\begin{figure}[t!]
\centering

\includegraphics[width=0.76\columnwidth]{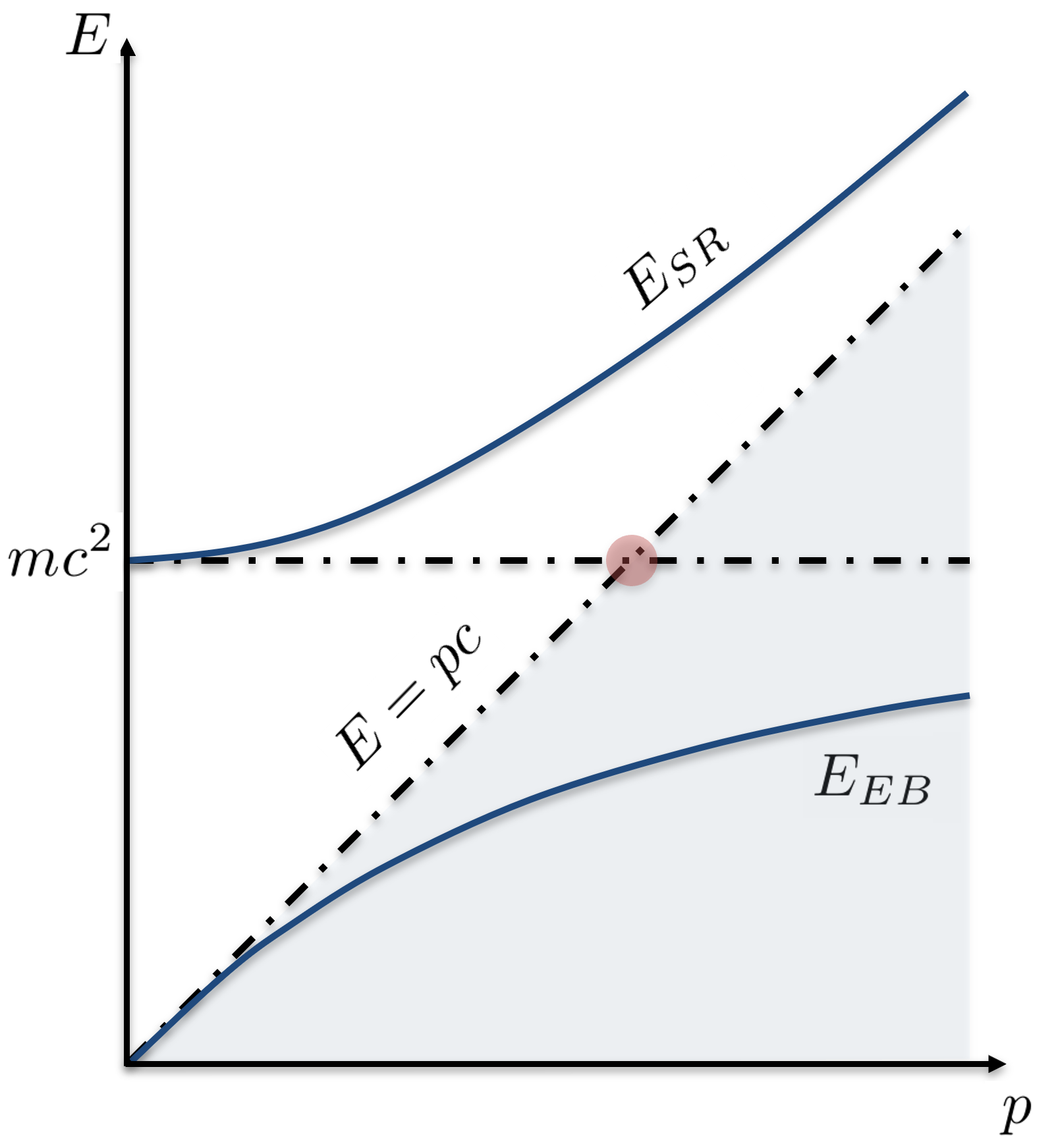}
\caption{The energy momentum relation (EMR) of a particle with constant rest energy (full line) and the EMR of a particle with pure kinetic energy $pc$ (dashed line) cross at the energy $E=mc^2$ (red dot). In our model the coupling of these states leads to an avoided crossing with the EMR of a free particle according to special relativity (dotted line) and the EMR of a particle in Einsteins basement (blue area; dashed-dotted line).}
\label{fig: Two branches}
\end{figure}

Our model starts with the notion of two conceptually different types of bare particles, namely those with a constant rest energy $E = mc^2$ and those with only kinetic energy $p c$ (e.g. light). Here, $E$  denotes the energy, $m$ the rest mass, and the $c$ the speed of light in vacuum, respectively. $p=\vert {\vec p} \vert$ is the modulus of the momentum of the particle \cite{ein05c, hec09}.  
We see from Fig. \ref{fig: Two branches} that these two EMRs (i.e. particles) have the same energy for $p = mc$, which means that their EMRs have an allowed crossing. However, for any coupling (interaction) between these particles the crossing will be avoided. It is a universal principle of physics - sometimes called {\bf no-level crossing theorem} - that a pair of energy levels connected by a perturbation do not cross as the strength of the perturbation is varied from zero \cite{sak21, neu29}.
Examples of avoided crossings are found everywhere in macroscopic physics and quantum mechanics, e.g. in mechanics \cite{nov10a, fau12, bro11}, optics \cite{hua51}, acoustics \cite{new17}, in quantum mechanics of molecules \cite{yan20a} or quantum chemistry \cite{nov10}, to name only a few. 
In the avoided crossing we have now two branches of the EMRs.

In our model we identify the upper branch with the well-known EMR of special relativity  
\begin{equation}
	E_{SR}(p) = \sqrt{(mc^2)^2 + {\vec{p} }^2c^2}.
\label{eq: E_1}
\end{equation}

This EMR is (the only one) intimately associated with the geometry of Minkowski spacetime governed by Lorentz symmetry. Its properties have been tested to its best over the last century. 
This EMR  is shown in Fig.~\ref{fig: Two branches} as the upper curve for the mass $m$. For low velocities it  starts with the rest energy $m c^2$ before it approaches a linear behavior for large kinetic energies. 
%\textcolor{red}{In quantum electrodynamics the linear dashed line represents the EMR of the photon. After the detection of %gravitational waves \cite{abb16x} this EMR could also be regarded as the graviton line. }

In a simple form, an avoided crossing can be described by a matrix $\cal{A}$ that couples the two bare states  with the coupling $F(p)$ as 
%\textcolor{green}{
\begin{equation}
 \cal{A}  =  
 \begin{pmatrix}
 p c & F(p, m)\\
  F(p, m) & mc^2
  \end{pmatrix}.
  \label{eq: Matrix}
\end{equation}
Diagonalizing this matrix leads to the dressed state $E(p)$ with the characteristic equation
%\begin{eqnarray}
%  \begin{vmatrix}
% p c - E(p) & F(p, m)\\
%  F(p, m) & mc^2 - E(p)
%  \end{vmatrix} \\
%= (p c - E(p)) (mc^2-E(p)) -F(p)^2 = 0. \nonumber 
%\end{eqnarray}
\begin{eqnarray}
\nonumber \\
E(p)^2 - (p c+mc^2)E(p)  + p c mc^2 - F(p,m)^2 =0 \,.
\end{eqnarray}
Assuming the first solution to be $E_{SR}(p) = \sqrt{p^2 c^2 + m^2c^4}$ the second solution is 
\begin{eqnarray}
E_{EB}(p) % &=& mc^2 + p c - E_{SR}(p) \nonumber \\ 
&=& mc^2 + p c -  \sqrt{(mc^2)^2 + {p}^2c^2}\,,
\label{eq: E_2}
\end{eqnarray}
where the subscripts $\it{SR}$ and $\it{EB}$ refer to special relativity and Einstein's basement, respectively \cite{remark_basement,Vieta2}.

The new EMR of Eq.~(\ref{eq: E_2}) is displayed also in Fig. \ref{fig: Two branches} for the same mass $m$.
For small momenta, the EMR approaches the $pc$-line whereas it clings to the rest mass for large momenta.   
With the coupling $F(p,m)$, the EMRs $E_{SR}$ and $E_{EB}$ represent quasi particles that can be regarded as the excitations of an (up to now) unknown 
field. 

In what follows, we will investigate how the kinematics of the particles can change when the physics of the lower branch becomes accessible. 

Having assumed, that energy is a valid concept in Einstein's basement - in the sense of a first integral of motion - we will also assume that Hamiltonian mechanics \cite{lan76} with the Hamiltonian ${\cal{H}} = {\cal{H}}({\vec{q}},{\vec{p}},t)$, and the canonical coordinates, position ${\vec{q}}$ and momentum ${\vec{p}}$, is suitable to describe the physics in both branches.

The velocity $\vec{v}$ of a particle is given by the Hamilton equation
\begin{equation}
\vec{v} = \dot{\vec{q}} = \nabla_{\vec{p}}\cal{H}\,.
\end{equation}
Hence, for ${\cal{H}} =E_{SR}$ we obtain
\begin{equation}
\vec{v}_{SR} = \frac{ p c^2}{\sqrt{(mc^2)^2 + {p}^2c^2}}  \hat{\vec{v}}\,,
\label{eq: velocity v1} 
\end{equation}
where $\hat{\vec{v}}$ is the velocity unit vector. In contrast, for the lower branch ${\cal{H}} =E_{EB}$ we get
\begin{eqnarray} 
\vec{v}_{EB}
&=& \left(c - \frac{p c^2}{\sqrt{(mc^2)^2 + p^2c^2}} \right) \hat{\vec{v}}\,.
\label{eq: velocity v2}
\end{eqnarray}
\begin{figure}[b!]
\centering
\includegraphics[width=0.96\columnwidth]{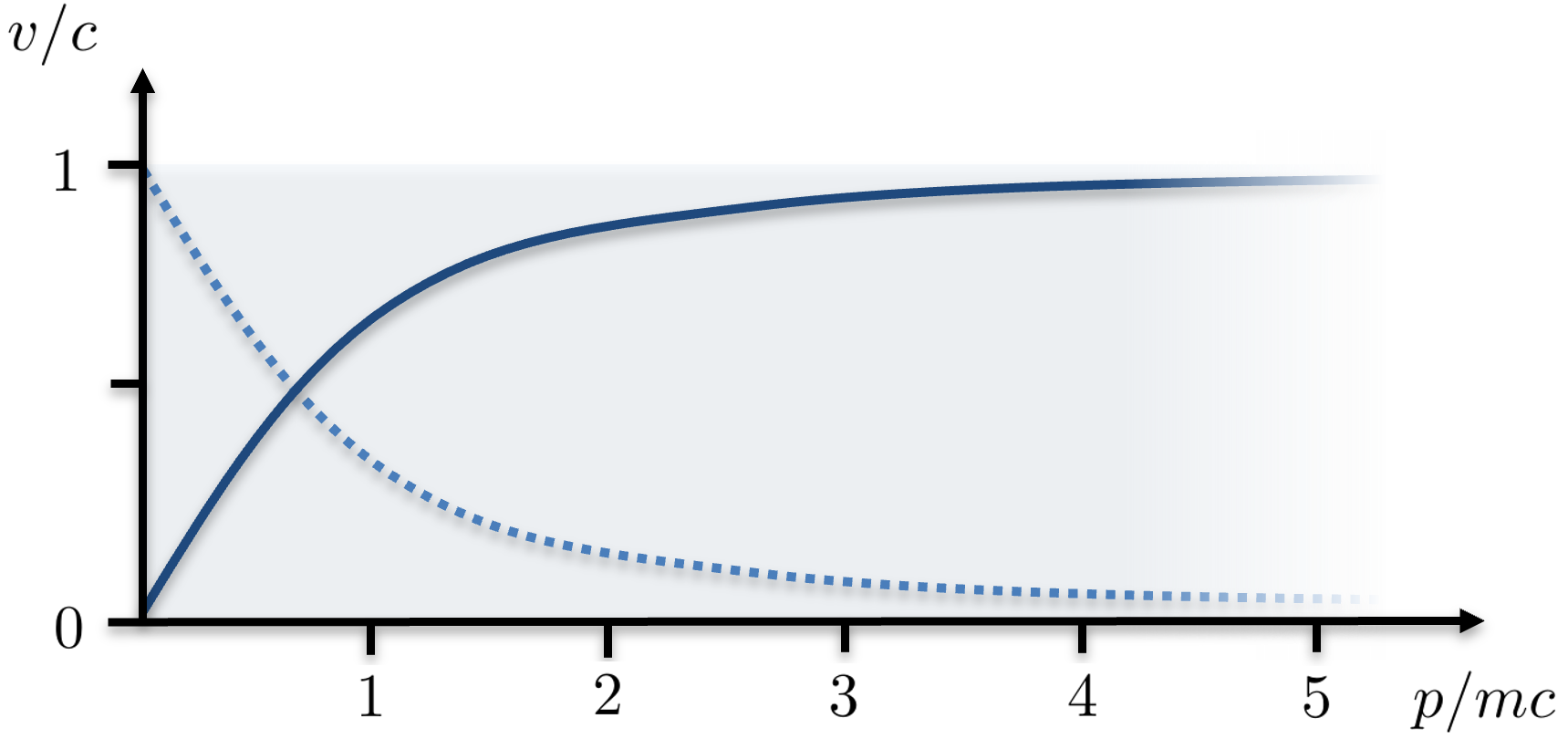}
\caption{Velocity of the particle as a function of the momentum $p$ according to Eq.~(\ref{eq: velocity v1}) and (\ref{eq: velocity v2}) for $E_{SR}$ (full line) and $E_{EB}$ (dashed line), respectively.}
\label{fig: velocity}
\end{figure}

Thus, we find that the velocity of a particle behaves differently in both cases.
In special relativity $|\vec{v}|=v$  depends linearly on the momentum for $v\ll c $ and approaches the velocity of light for asymptotically high momenta.
For particles in Einstein's basement $v$ decreases from $c$ for low momenta and goes to zero for $p\to \infty$.
Both behaviors are illustrated in Fig. \ref{fig: velocity}.

From Eq.~(\ref{eq: velocity v1}) one derives the well known momentum of a relativistic particle 
\begin{equation}
\vec{p}_{SR} = \frac{mv}{\sqrt{1-\left( \frac {v}{c} \right)^2}} \hat{\vec{v}}.
\label{eq: momentum1}
\end{equation}
Doing the same for the lower branch, we obtain 
\begin{equation}
\vec{p}_{EB}  =  \frac{m(c-v)}{\sqrt{1- (\frac{c-v}{c})^2}} \hat{\vec{v}}.
\label{eq: momentum2}
\end{equation}
In Einstein's basement, the well known {\it Lorentz factor} of special relativity
\begin{equation}
\gamma_{SR} = \frac{1}{\sqrt{1- (\frac{v}{c})^2}} 
\label{eq: Lorentz factor}
\end{equation}
is replaced by a corresponding factor 
\begin{equation}
\gamma_{EB} = \frac{1}{\sqrt{1- (\frac{c-v}{c})^2}}.
\label{eq: Basement L factor}
\end{equation}
We find that both momenta have a similar form, while in Eq.~(\ref{eq: momentum2}) the velocity $v$ is replaced by $c-v$ in comparison to the relativistic case \cite{Symmetrie}.

\begin{figure}[tbh]
	\centering
	\includegraphics[width=0.7\columnwidth]{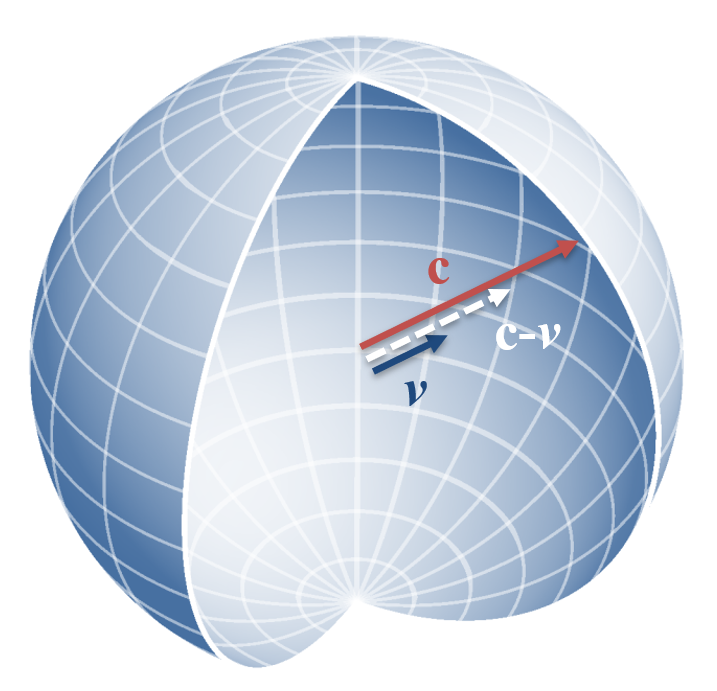}
	\caption{In velocity space the constant speed of light is visualized by a sphere where a particular velocity $\vec{v}$ and the corresponding $\vec{c}-\vec{v}$ are depicted. We see that for each velocity $\vec v$ one can find a parallel vector ${\vec c}$ to construct $\vec c-\vec v$ with a magnitude $c-v\in [0,c]$, used in Eqs.~(\ref{eq: momentum2}) and (\ref{eq: Basement L factor}).}\label{fig: c-Kugel}
\end{figure}

Hence, both, (\ref{eq: Lorentz factor}) and (\ref{eq: Basement L factor}) are linked by a mirror symmetry \cite{Symmetrie} 
 \begin{equation}
v \leftrightarrow c -v 
\label{eq: VauStrich} 
\end{equation}
which is visualized in Fig. \ref{fig: c-Kugel}.
Thus, we observe that both branches respect $c$ as an ultimate limit for the motion of particles, while the roles of $v=0$ and $v=c$ are interchanged.
In the relativistic case, the special relation of the momentum (\ref{eq: momentum1}) and the energy (\ref{eq: E_1}), gives rise to the class of Lorentz transformations, keeping $\sqrt{E_{SR}^2-c^2\vec{p}_{SR}^2}=mc^2$ invariant. However, this does not hold for the lower branch, where no such transformation can be found. In particular, $\sqrt{E_{EB}^2-c^2\vec{p}_{EB}^2}$ is not Lorentz invariant by itself, i.e., has a different value in every Lorentz frame, giving up this principle of relativity in the lower branch. 
We will see that this feature will also translate into the equations of motion, when we investigate the dynamics of the newly introduced quasi particles. 

%#################################################################

Group theory \cite{hal15c} and differential geometry \cite{nee21d} allow us to get deeper insights in the framework behind special relativity and Einstein's basement. 
With the mirror transformation Eq.~(\ref{eq: VauStrich})
(i.e. an opposite isometry) between the dynamics of both sectors, an overarching symmetry group must complement the Poincar{\'e} group by shifts and reflections in velocity space.
The full elaboration of these thoughts in group theory will not be treated here, since here we are primarily interested in the novel kinematics in Einstein's basement to be discussed in Sec.~ \ref{sec:new_dynamics}.
But before, we will discuss a quantum optical system in solid state physics that already displays the same essential properties as our model does.

\section{Polaritons - a quantum optical system equivalent to our model}
\label{sec: Polaritons}
\begin{figure}[tbh]
	\centering	\includegraphics[width=0.76\columnwidth]{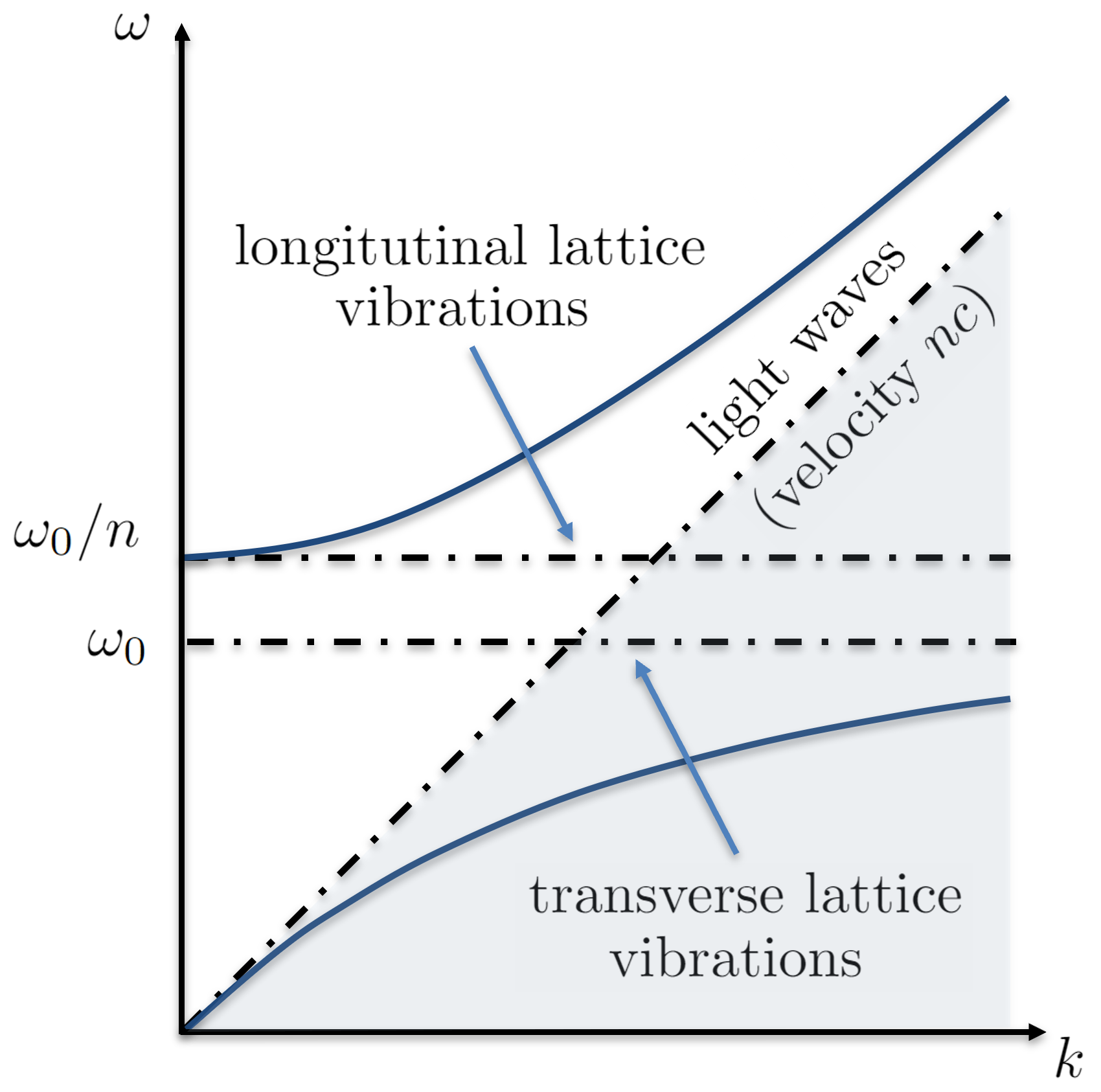}
	\caption{
		The energy momentum relationships of phonon-polaritons after \cite{hua51} (see text).}
	\label{fig: polaritons}
\end{figure}

An equivalent system to Einstein's basement can be found e.g. in the (phonon)-polaritons in an ionic crystal \cite{kit05, hua51, hua51a} or in the coupled magnon-phonon modes in a ferromagnet in the Faraday configuration \cite{mil74r}. The phonon-polariton system shows the typical dispersion relations of Fig. \ref{fig: polaritons}. 
%------------------
In the ionic crystal usually two mechanical eigenfrequencies  can be excited that correspond to transverse and longitudinal oscillations (phonon modes) \cite{Polariton} of the atomic layers \cite{kit05}. An electromagnetic wave can interact with the ions via its electric field. The EMRs of this system (Fig. \ref{fig: polaritons}) resemble strikingly the EMRs of Fig. \ref {fig: Two branches}, which is not surprising since one starts with the same bare states of light and modes with constant energies that we have used to construct our model.  
In this system we find a gap resulting from the different frequencies of the longitudinal and transversal modes. This gap is not present in model (Fig. \ref{fig: Two branches}) as it is also the case in some polariton systems, such as in germanium or silicon crystals \cite{kit05}.
%---------------------------
Observers in this system who are aware of the upper branch only, would conclude (like we do in special relativity) that Lorentz transformations are valid everywhere and that the physical laws have to be the same in each inertial system. If they would find the lower branch, however, these observers would conclude that both sorts of EMRs i.e. both sorts of quasi particles can persist in the same physical system as we investigate in this paper. 

Such a ``pseudo-relativistic behavior" of the upper EMR 
of a polariton system
has also been observed in other crystalline systems \cite{unz00,gue96}.
The formally identical conditions between our model and the polariton system in a crystal could be used as a toy model  for experiments to visualize special properties of particles in Einstein's basement and their interaction \cite{tar00}.

%#################################################################

\section{New dynamics for particles from Einstein's basement} 
\label{sec:new_dynamics}

To study the kinematics of particles of Einstein's basement, we want to construct the counterpart of the relativistic Lagrangian 
\begin{eqnarray}
\mathcal{L}_{SR}&=& - m c^2 \sqrt{1-\left(\frac{v}{c}\right)^2}\label{eq: Lagrangian 1}\\
&=& -m c^2 +\frac{m}{2}\, v^2- \dots\,.
\nonumber
\end{eqnarray} 
for a free particle, where an expansion for small velocities $v\ll c$ reveals the well-known Newtonian kinetic energy term.
Keeping in mind that the relation
\begin{equation}
\vec{p}_{EB} = \frac{\partial {\cal L}_{EB}}{\partial \vec{v}}
\end{equation}
holds, we integrate  Eq.~(\ref{eq: momentum2}) to obtain the Lagrangian
\begin{eqnarray}
{\cal L}_{EB} &=& - {m} c^2 + {m} c^2 \sqrt{1 - \left(\frac{c-v}{c}\right)^2}\label{eq: Lagrangean2}\\
&=& -{m}c^2+\frac{{m}}{2}(2c)^{3/2}\sqrt{v}-\dots
\nonumber
\end{eqnarray}
for a free particle of Einstein's basement, where
the integration constant was chosen as $-{m}c^2$ to obtain the Lagrangian for a particle at rest. 
In analogy to the second line of Eq.~(\ref{eq: Lagrangian 1}), we obtain a kinetic energy term in the low-velocity limit of Eq.~(\ref{eq: Lagrangean2}). Remarkably, also in this limit, the speed of light $c$ remains as a parameter in the system, affecting the motion of such particles.
In the next section, we will analyze this expression and discuss the resulting dynamics of particles, which qualitatively differs from the Newtonian case in many aspects.

In the following, we have a closer look at the dynamics in Einstein's basement by using the free particle Lagrangian Eq.~(\ref{eq: Lagrangean2}) in the low-velocity limit and complement it by an additional potential depending on the particle's position in space
\begin{equation}
\tilde{ \mathcal{L}}=\mathcal{L}_{EB}-V(\vec{q})\,.
\end{equation}
The corresponding Euler-Lagrange equations 
\begin{equation}
\frac{d}{dt} \frac{\partial {\cal L}_{EB}}{\partial\vec{v}}= -\frac{\partial V(\vec{q})}{\partial {\vec{q}}}
\label{eq: EulerLagrange}
\end{equation}
lead to the equations of motion in the form
\begin{equation}
   m \,\vec{a} -\frac{3  m (\vec{a}\cdot\vec{v})}{2 v^2} \,\vec{v}=\frac{1}{2} \left(\frac{2 v}{c}\right)^{3/2} \vec{F} \,,\label{eqn:newton_modifyed_1}
\end{equation}
where $\vec{a}=\ddot {\vec q}$ is the particle's acceleration and $\vec{F}$ the force resulting from the potential. 
We find that the time derivative of the momentum $\frac{d}{dt}\vec{p}_{EB}\not= m\, \vec{a}$, does not coincide with the acceleration, as it is the case in Newtonian physics. 
In contrast, $ m\, \vec{a}$ is complemented by an additional term, which depends on the particle's velocity $\vec{v}$. By rearranging Eq.~(\ref{eqn:newton_modifyed_1}) into the form \begin{equation}
     {m} \,{\vec{a}} =\frac{1}{2}  \left( \frac{2v}{c} \right)^{3/2} \left(\vec{F}-3\frac{(\vec{F}\cdot\vec{v})}{v^2}\,\vec{v}\right)
\end{equation}
the equations of motion can be directly compared to Newton's second law. For $\vec{F}=0$ we obtain the result $\vec{a}=0$ as in the Newtonian case, leading to a uniform motion of the particle.
In the presence of external forces, however, the dynamics of the particle 
differs from Newton's second law in several aspects. 
We find that the effect of the external force is drastically reduced for slow particles due to the overall-prefactor, depending on  $|\vec{v}|=v \ll c$. However, the orientation of $\vec{v}$ with respect to $\vec{F}$  
affects the term in the second brackets. Due to that, the particle's response to the force can be positive or negative, depending on the velocity vector.
 For instance, in the special case when the force and the velocity are parallel (or anti-parallel), we obtain an overall negative sign
\begin{equation}
 {m} \,{\vec{a}} = - \left( \frac{2v}{c} \right)^{3/2}  \vec{F}\,,
\label{eq: Newton_2_basement}
\end{equation}
such that in this case commonly attractive forces, act repulsive on a ``basement particle''.
We stress, that this repulsive force has its origin in the particle dynamics and must not be confused with the assumption of a negative particle mass, as it was assumed in dark energy considerations by A.~Einstein \cite{ein18c}, or more recently by  J.~S.~Farnes \cite{far18}.

Having disclosed the unconventional behavior of particles in the lower branch of Fig.~\ref{fig: Two branches}
with respect to external forces, it is straightforward to investigate the 
interaction of basement particles with each other 
and with common matter.
In the following, we will consider a purely gravitational interaction between two classical particles, two particles in Einstein's basement, and between one of each kind.

\subsection{Dynamics of two classical particles}
In order to study the gravitational interaction of the newly introduced particles, we first recall the central features of the Keplerian motion of two regular particles of masses $m_1$ and $m_2$. The corresponding Lagrangian is given by
\begin{equation}
\mathcal{L}_{SR-SR}=\frac{m_1}{2}v_1^2+\frac{m_2}{2}v_2^2+\frac{Gm_1 m_2}{|\vec{q}_1-\vec{q}_2|}
\end{equation}
By utilizing the Euler-Lagrange equations and the definition of the relative and center of mass coordinates
\begin{eqnarray}
    \vec{r}&=&\vec{q}_1-\vec{q}_2\\
    \vec{s}&=&\frac{m_2}{m_1+m_2}{\vec{q}}_2+\frac{m_1}{m_1+m_2}{\vec{q}}_1\,,
\end{eqnarray}
the dynamics of the system is described by
\begin{eqnarray}
\ddot{\vec{r}}&=&-\frac{G(m_1+m_2)}{|\vec{r}|^3}\,\, \vec{r}\label{eq: Kepler}\\
\ddot{\vec{s}}&=&0\,. \label{eqn:COM}
\end{eqnarray}
Therefore, one finds that the relative motion of the particles and their center of mass motion can be described independently.  The equation~(\ref{eq: Kepler}) leads to the well-known Kepler orbits while the particle's center of mass moves with a constant velocity.
We will see that neither the first nor the second condition will hold when particles from Einstein's basement are involved.
%-------------------------------
\subsection{Dynamics of two slow basement particles}
In analogy to the latter case we again define the Lagrangian
\begin{eqnarray}
\mathcal{L}_{EB - EB} &=&\frac{m_1}{2}(2c)^{3/2}\sqrt{v_1}+\frac{m_2}{2}(2c)^{3/2}\sqrt{v_2} \nonumber \\
&+&\frac{Gm_1 m_2}{|\vec{r}|} 
\end{eqnarray}
of two gravitationally interacting particles using Eq.~(\ref{eq: Lagrangean2}) in the low-velocity limit.
The equation of relative motion for the two particles reads
\begin{eqnarray}
    \ddot{\vec{r}}&=&-\frac{\sqrt{2} G\left(m_1 (v_2/c)^{3/2}+m_2 (v_1/c)^{3/2}\right)}{|\vec{r}|^3}\,\, \vec{r}
    \nonumber \label{eq: 2BasementParticles} \\
    & &+\,\frac{3\sqrt{2} G}{c^{3/2} |\vec{r}|^3}\left(\frac{ m_2  ( \vec{v}_1\cdot\vec{r})}{\sqrt{v_1}}\,\,\vec{v}_1+\frac{ m_1  ( \vec{v}_2\cdot\vec{r})}{\sqrt{v_2}}\,\,\vec{v}_2\right)\!.\,\,\, \label{eqn:EB-EB}
  \end{eqnarray}
This equation contains a number of important and counter-intuitive features. First, it shows that the motion of the particles does not only depend on their relative position $\vec{r}$, but also on their velocities $\vec{v}_1$ and $\vec{v}_2$ with respect to an absolute space. Second, the first term of Eq.~(\ref{eq: 2BasementParticles}) is attractive like in the Newtonian case from Eq.~(\ref{eq: Kepler}), but with a velocity-dependent mass term $\sqrt{2}[m_1 (v_1/c)^{3/2}+m_2 (v_2/c)^{3/2}]$ instead of the total mass $m_1+m_2$ of the system.
Thus, for small velocities $v_i\ll c$ the 
mass effectively entering the gravitational interaction is drastically reduced in comparison to the classical two-body problem. 
Third, 
we find the second, also velocity-dependent  term in Eq.~(\ref{eq: 2BasementParticles}) which is absent in the Newtonian case. It is always repulsive and can be up to three times larger than the attractive term. 
Consequently, the interplay of these two contributions decides whether the net acceleration between the two particles is positive or negative. 

\begin{figure}[t!]
\centering
\includegraphics[width=\columnwidth]{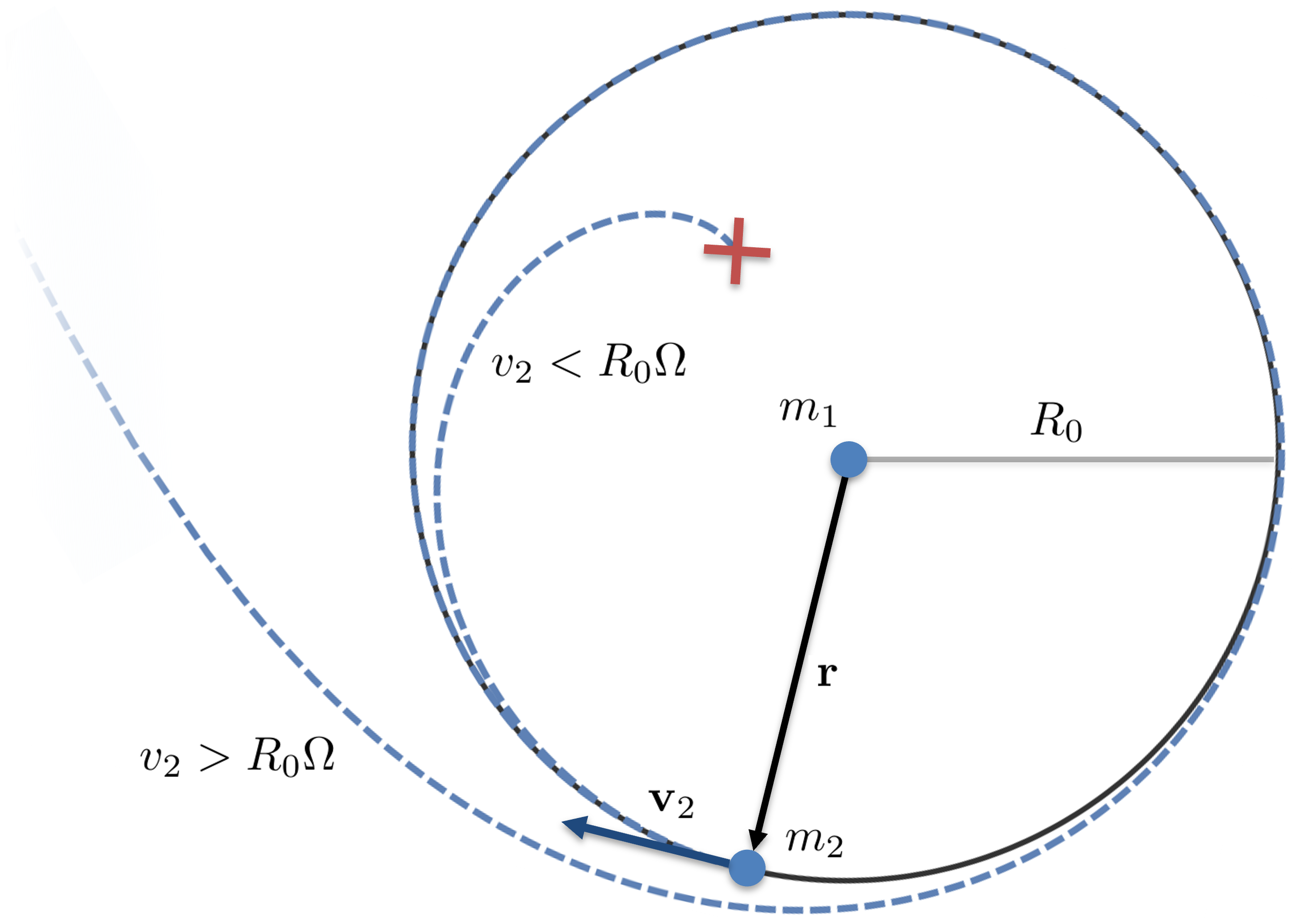}
\caption{Orbit of a basement particle with mass $m_2$ and velocity $\vec{v}_2$, orbiting another basement particle of mass $m_1$ at rest.
According to Eq.~(\ref{eq: 2BasementParticles}) the moving particle describes a circular orbit
in the particular case of $v_2=R_0\Omega$.
Below this value, for $v_2<R_0\Omega$, the second particle comes to a final stop (at the red cross) within the circular orbit.
For $v_2>R_0\Omega$ the second particle increases its distance and eventually escapes to infinity.}
\label{Fig3}
\end{figure}

To study the dynamics of two gravitationally interacting basement particles in more detail we consider the three ultimate cases in Fig. \ref{Fig3}.
For that purpose we assume the first particle to be at rest with respect to absolute space, i.e., $\vec{v}_1=0$. 
Further assuming $\vec{v}_2\cdot\vec{r}=0$, we find that there exists
a circular orbit of the second particle with $|\vec{r}|= R_0= \mathrm{const}$ and a constant frequency $\Omega =\frac{2 G^2 m_1^2}{c^3 R_0^3}$. 
The motion of the second particle, regardless of its own mass $m_2$, only depends 
on the mass $m_1$ of the fixed particle and the distance $R_0$.
This follows naturally recalling that the momentum of a basement particle with zero velocity is infinite, cf. Eq.~(\ref{eq: velocity v2}) and Fig.~\ref{fig: velocity}.
The circular orbits with velocity $v_2=R_0 \Omega$ are unstable and mark the transition between two regimes of distinct behaviour.
If the velocity $v_2 > R_0 \Omega$ of the second particle is increased by a very small amount, the particle escapes to infinity, as shown in Fig.~\ref{Fig3}b. 
If the velocity $v_2 < R_0 \Omega$ of the second particle is reduced, the particle  turns into the region of the circular area and ultimately comes to rest at a finite distance, as shown in Fig.~\ref{Fig3}c.
Such a stopping point is also encountered when two basement particles of arbitrary mass move towards each other on a straight line in absolute space. 
%-----------------------------------
\subsection{Dynamics of a classical and a (slow) basement particle}
\label{sec:SR-EB}
\begin{figure}[htb]
\centering
\includegraphics[width=0.8\columnwidth]{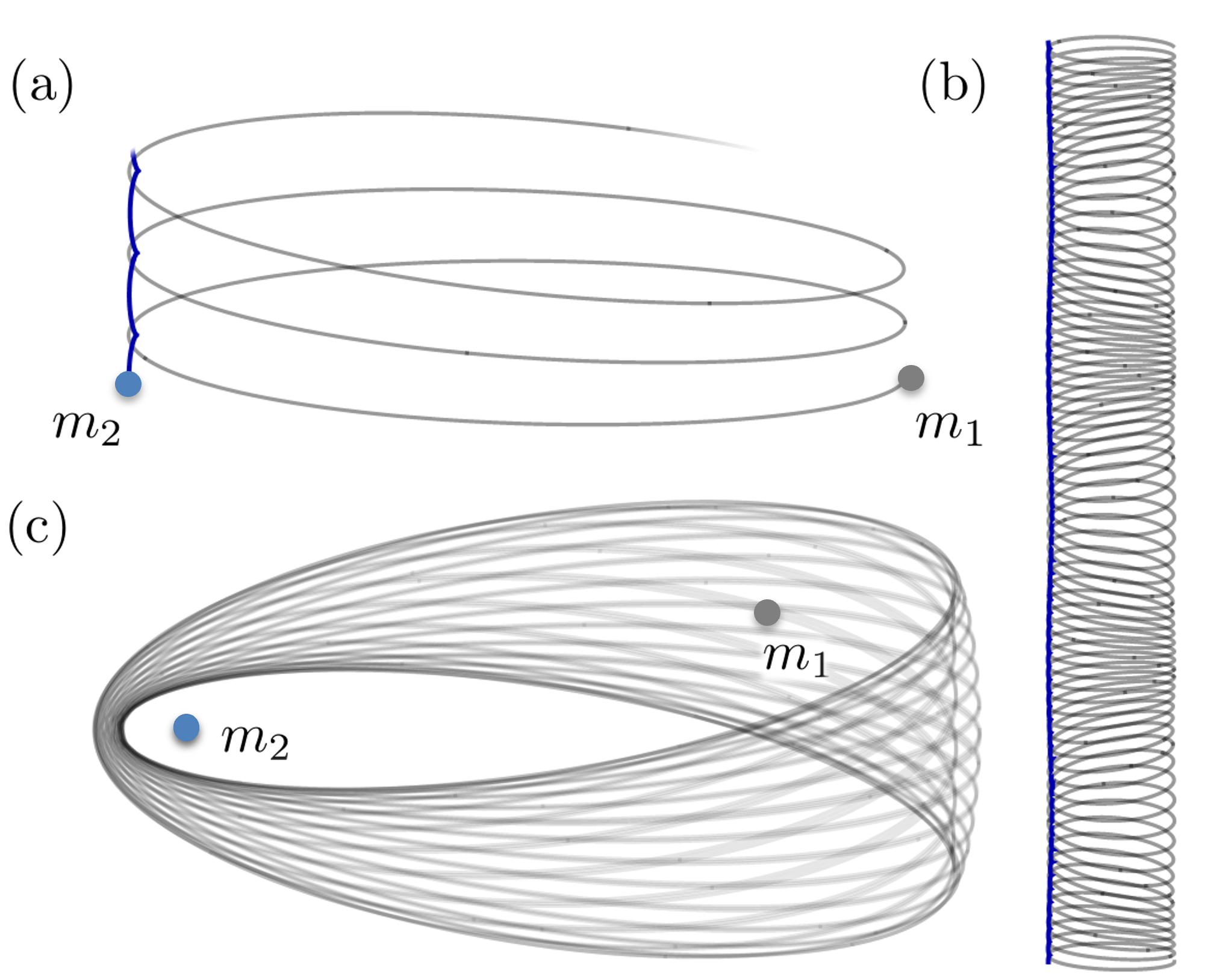}
\caption{Simulated trajectories of a regular particle with mass $m_1$ (black) and a slow basement particle with mass $m_2$ (blue), interacting gravitationally. 
The basement particle is almost unaffected by its companion, whereas
the regular particle follows the former one in a spiraling motion, see (a) and (b).
The relative orbit of $m_1$ with respect to $m_2$, given in (c), shows two periodicities: an elliptic orbit comparable to Kepler motion, and an oscillation of the apsidal position \cite{num_parameters}.} 
\label{Fig4_Mertens}
\end{figure}

Here, we want to discuss the behavior of the relative motion and the non-conservation of the center of mass velocity for the gravitational interaction of classical and a basement particle. Beginning with the Lagrangian
\[
\mathcal{L}_{SR - EB}=\frac{m_1}{2}v_1^2+\frac{m_2}{2}(2c)^{3/2}\sqrt{v_2}+\frac{Gm_1 m_2}{|\vec{r}|}
\]
and the corresponding Euler-Lagrange equations we find
\begin{eqnarray}
	\ddot{\vec{q}}_1&=&-\frac{Gm_2}{|\vec{r}|^3} \vec{r} \label{eqn:CP-EB01}\\
\ddot{\vec{q}}_2&=&\frac{\sqrt{2} G m_1\left( v_2^2 \,\vec{r}-3 ( \vec{v}_2\cdot\vec{r})\,\vec{v}_2
\right)}{c^{3/2} |\vec{r}|^3 \sqrt{v_2}}\,. \label{eqn:CP-EB02}
\end{eqnarray}
 Other than in the Lagrangian, at the level of the equations of motion we can now easily perform the limit $v_2/c\to 0$ and find 
\begin{eqnarray}
	\ddot{\vec{r}}=-\frac{G m_2}{|\vec{r}|^3} \vec{r}\qquad&,&\qquad 
	\ddot{\vec{q}}_2=0\,, \label{eqn:test_particle_case}
\end{eqnarray}
where $\ddot{\vec{r}}=\ddot{\vec{q}}_1$ holds in this case.
We find, that the classical particle behaves like a test particle attracted by the gravitational potential of a fixed particle of mass $m_2$. Within this approximation the basement particle is still allowed to move with a constant but small velocity $v_2 \ll c$, such that the total system  acts like a classical system with $m_1 \ll m_2$ under Galilean transformation. Other than that, however, the relation between the masses plays no role if the particle of mass $m_2$ is a slow basement particle. 
%Moreover, a basement particle is repelled from a regular particle whereas the latter one is fully attracted by the basement particle. For two masses $m_2 \approx m_1$ the repulsion is extremely weak as compared to the attraction. \remark{[?!]} 
We see from Fig. \ref{Fig4_Mertens} that the motion of the particle in Einstein's basement is virtually not affected by the classical particle whereas the latter one's motion encircles the first one. Hence, the center of mass velocity of the combined system is not a constant of motion. 

\section{Summary, Conclusions and Outlook}
\label{sec:conclusion}
We have complemented the EMR$_{SR}$ of special relativity by a second EMR$_{EB}$ in Einstein's basement derived from a forbidden crossing between the constant rest mass of a particle and a particle with energy $pc$. As a result, regular matter and matter originating from Einstein's basement are considered as made of quasi particles of a yet unknown field. We studied the kinematics of the newly introduced particles and found that their velocity $v$ is replaced by $c-v$ in comparison to the regular case.
\begin{comment}
Our findings might possibly be tested quantitatively at different scales, for instance by astrophysical observations of stars, nebulae and galaxies
and by gravitational lensing effects of regular and dark matter components. 
A promising candidate for such an investigation could be the so-called bullet cluster \cite{par16a}.
This cluster comprises of two components  each with a spearhead dark matter contribution followed by a distribution of hot regular gaseous material.
In such a geometry our model asks - in contrast to regular kinematics - for additional acceleration of the two dark matter components away from each other, which might be observable in successive measurements.

It must be checked, whether our model
delivers an competitive explanation for the existing measurement data and provides an alternative to the current maps of the dark matter distribution in the galaxies \cite{spr05, hon21}.
Revisiting the huge number of existing data and the new data expected from the EUCLID space mission \cite{EUCLID} could give a hint whether the presented ideas should be pursued further.

If our model can survive the comparison with astrophysical observations there would be many more options to explore new physics.
\end{comment}
In this paper, we only considered classical particle dynamics and compared our results to those of classical Newtonian physics.
However, with the full Lagrangian (\ref{eq: Lagrangean2}) at hand, 
an extension of our analysis towards a comparison with special and general relativity is possible.
This is of particular interest, since the concept of relativity does not hold for the sector of Einstein's basement, while it is still preserved for regular matter in our model. 
Considering interactions between particles of both sectors, violations of the relativity principle, such as a breaking  of local Lorentz invariance, can be expected also for regular matter. 
In theory, extensions of the standard model of particles have been proposed which include general relativity and many possible operators that break Lorentz symmetry \cite{col98}.
Investigations of such potential violations are ongoing both theoretically \cite{kos03} and experimentally with different means in pulsars \cite{sha14, don24}, ions \cite{meg19}, atom interferometers \cite{chu09} and others.

Assuming that Newtonian gravity also holds for the quasi particles in Einstein's basement, we found that under certain conditions also repulsion between the particles can occur. Furthermore, regular particles are dragged along with basement particles. 
Combining these two findings, this mechanism can describe an expansion of a mixed system {\it in space}. Whether this hypothesis can contribute or explain the observed accelerated expansion of the universe, as it is suggested by the study of type Ia supernovae candles \cite{sch98a,per99}, needs to be investigated.

Einstein's basement might also be a region where the long searched for constituents of dark matter could hide. Properties of dark matter particles have been investigated \cite{ber05, ber18} and a number of theoretical hypotheses has been offered for the origin of dark particles like weakly interacting massive particles (WIMPs) \cite{jun96}, axion-like particles \cite{kam97}, dark photons \cite{fab20x}, erebons \cite{pen18}, Kaluza-Klein particles \cite{col20} or others.
However, none of these particles have been found experimentally up to now \cite{zha22x, apr22}. A possible reason might be that the searched particles may obey a different kinematics, such as the one described in our model. 

The existence of two branches of different energies (Fig. \ref{fig: Two branches}) suggests {\it diabatic} transitions under a suitable interaction between the two branches. Diabatic electromagnetic transitions are well known to occur in the case of an avoided crossing. Such transitions were modeled by Landau \cite{lan32}, Zener \cite{zen32}, St{\"u}ckelberg \cite{stu32} and Majorana \cite{maj32} and have been tested experimentally in quantum optics \cite{rub81}. Assuming that transitions under the influence of a particular interaction are allowed, they would lead to a decrease of the population in the $E_{SR}$ state and increase the population in the $E_{EB}$ state. However, since no signals of such transitions have been observed so far, we conclude a very low probability for the decays to occur and can exclude electroweak and strong interaction as initiating such a decay. 
Considering interactions of this kind unlikely, particles from both sectors may be still allowed to interact gravitationally 
by virtue of their positive energy densities.
%Under the rough assumption that at the beginning of the universe all particles were in the upper state and now after $3.7 \times 10^{10}$ years 80\% are as dark matter in the lower state we find a very low probability for the decays to occur and can exclude electroweak and strong interaction as initiating such a decay. 

A future task is to checked, whether our model
delivers an competitive explanation for the existing measurement data and provides an alternative to the current maps of the dark matter distribution in the galaxies \cite{spr05, hon21}. Revisiting the huge number of existing data and the new data expected from the EUCLID space mission \cite{EUCLID} could give a hint whether the presented ideas should be pursued further. There is furthermore to be checked 
whether the idea of Einstein’s basement can be applied at macroscopic, i.e., astrophysical or cosmological scales or whether it can provide meaningful interpretations only at the scales of quantum physics, such as it is the case for Dirac theory.

A field theoretical approach of our model could elucidate the nature of the field whose excitations are the quasi particles, we identified as regular particles and particles from Einstein's basement. 
If we stress the similarities of our model with the description of polaritons in a crystal \cite{kit05, hua51, hua51a}, the underlying field would correspond to the crystal lattice in that case.
In particular, the relation of the mass or energy density of this hypothetical background field to concepts like the cosmological constant and dark energy might be a promising investigation.

A consequent extension of the work presented here is to study the given EMRs in a quantum mechanical framework and, ultimately, within the formalism of quantum field theory.
Since the inner symmetries of the fields, such as the $U(1)$-gauge symmetry,
exist independently of the considered group of motion, it is straightforward to write a theory of gauge fields, either in special relativity or in Einstein's basement. 

Particularly interesting is the regime of small energies and momenta, where all EMRs in Einstein's basement closely approach each other near the $pc$-line.

\begin{acknowledgments}
We thank Uwe Sterr, Piet Schmidt, Andrey Surzhykov, Eberhard Tiemann and Jun Ye for fruitful discussions and Christopher Mertens for numerical simulations and Hauke Uhde for his analysis of the implications of our model for galaxy dynamics (work in progress). We acknowledge the support by the Deutsche Forschungsgemeinschaft (DFG, German Research Foundation) under Germany’s Excellence Strategy—EXC 2123 QuantumFrontiers—390837967. FR thanks QUEST Leibniz Research School for longstanding support.
\end{acknowledgments}

%% to avoid duplicate bib-files, please set the windows environment variables: 
%% BIBINPUTS = O:\4-3\4-3-Alle\Papers\TeXBib
%BIBINPUTS = D:\TeXBib
%\bibliography{D:TEXBib\texbi431,Riehle2,er,  Regular matter and dark matter have to be regarded as ``quasi particles''plement_ref}
%\bibliography{D:/TEXBib/texBi431,G:/EinsteinsBasement/Riehle2,G:/EinsteinsBasement/supplement_ref}

%Für den Rechner zu Hause und von USB-Stick---------------------------------
%\bibliography{D:/TEXBib/texBi431, E:/EinsteinsBasement/PRL_V2/Riehle2}
%-------------------------------------------------------------------------------------
%\bibliography{O:/4-3/4-3-Alle/Papers/TeXBib/texbi431,C:/Users/Riehle01/Desktop/EinsteinsBasement/PRL_V2/Riehle2}

%Für den Rechner im Büro und von USB-Stick (in F:)---------------------------------------------------------
%\bibliography{O:/4-3/4-3-Alle/Papers/TeXBib/texbi431,E:/EinsteinsBasement/PRL_V2/Riehle2} 
%-------------------------------------------------------------------------------------------------------------------

%Für den Rechner als Standalone System ---------------------------------------------------------
\bibliography{Riehle2, texbi431} 

\begin{thebibliography}{59}
\bibitem [{rem()}]{remark_basement}%
  \BibitemOpen
  \href@noop {} {}\bibinfo {note} {We have to point out clearly that the phrase ``Einstein's basement" refers only to the lower branch EMR$_{EB}$ and by no means to a special region of position space, which is populated by both sorts quasi particles.}\BibitemShut {Stop}%
\bibitem [{Vie()}]{Vieta2}%
  \BibitemOpen
  \href@noop {} {}\bibinfo {note} {We may also derive the coupling $F(p, m) = \left[ p c\cdot mc^2\!-\!\sqrt{p^2 c^2 + m^2c^4} \left(mc^2 \!+ p c - \!\sqrt{p^2 c^2 + m^2c^4}\right) \right]^{1/2}$.}\BibitemShut {Stop}%
\bibitem [{Sym()}]{Symmetrie}%
  \BibitemOpen
  \href@noop {} {}\bibinfo {note} {It is very likely that starting with the discrete symmetry operation $v \leftrightarrow c-v$ would allow to establish an overarching model for both, special relativity and Einstein's basement.}\BibitemShut {Stop}%
\bibitem [{Pol()}]{Polariton}%
  \BibitemOpen
  \href@noop {} {}\bibinfo {note} {In a typical ionic crystal the EMRs of the polaritons usually have more than one lower branch that lead to a gap between the different lower branches. However, in crystals of C, Ge, Si there is only one.}\BibitemShut {Stop}%
\end{thebibliography}
%-------------------------------------------------------------------------------------------------------------------

%G:/EinsteinsBasement/supplement_ref}
%\bibliography {C:/DataRiehle/TeXBib/texBi431,C:/Users/Riehle01/Desktop/EinsteinsBasement/Riehle2,C:/Users/Riehle01/Desktop//EinsteinsBasement/supplement_ref}
\bibstyle{osa}
%\bibliography{texbi431,supplement_ref}
%\bibliographystyle{unsrt}
%\bibliography{sample}
%\input{test.bbl}
\end{document}